\documentstyle[aps,jpc,preprint]{revtex}
\tightenlines
\begin{document}


\title{Quasi-chemical Theories of Associated Liquids}


\author{Lawrence R. Pratt }
\address{ Theoretical Division, Los Alamos National Laboratory, Los
Alamos, New Mexico 87545 USA }
\author{Randall A. LaViolette}
\address{Idaho National Engineering and Environmental Laboratory, PO
Box 1625, Idaho Falls, ID 83415-2208 USA}

\author{LA-UR-98-991}

\date{\today}

\maketitle

\begin{abstract}
It is shown how traditional development of theories of fluids based
upon the concept of physical clustering can be adapted to an
alternative local clustering definition.  The alternative clustering
definition can preserve a detailed valence description of the
interactions between a solution species and its near-neighbors, {\it
i.e.,\/} cooperativity and saturation of coordination for strong
association.  These clusters remain finite even for condensed phases.
The simplest theory to which these developments lead is analogous to
quasi-chemical theories of cooperative phenomena.  The present
quasi-chemical theories require additional consideration of packing
issues because they don't impose lattice discretizations on the
continuous problem.  These quasi-chemical theories do not require pair
decomposable interaction potential energy models.  Since calculations
may be required only for moderately sized clusters, we suggest that
these quasi-chemical theories could be implemented with computational
tools of current molecular electronic structure theory.  This can
avoid an intermediate step of approximate force field generation.
\end{abstract}
\pagebreak



\section{Introduction}
Recent molecular calculations\cite{Hummer:97c,Martin:97} have
suggested in new ways that a chemical perspective can be helpful in
computing thermodynamic properties of water and aqueous solutions.
This paper follows-up those observations and develops quasi-chemical
theories for the thermodynamics of associated liquids. 

The most direct antecedent of this effort was the recent calculation
of the absolute hydration free energy of the ferric ion (Fe$^{3+}$)
\cite{Martin:97}.  That calculation used electronic
structure results on the Fe(H$_2$O)$_6$$^{3+}$ cluster and a simple,
physical estimate of further solvation effects.  Those results were
organized according to the pattern of a simple chemical reaction and a
surprisingly accurate evaluation of the hydration free energy was
obtained.  A second important antecedent of this work was the recent
calculation of the free energy contribution due to electrostatic
interactions in liquid water\cite{Hummer:97c}.  That work
systematically exploited the observed distribution of close neighbors
of a water molecule, the distribution of {\it clusters,\/} in liquid
water in order to obtain an accurate, simple treatment of the
thermodynamics of electrostatic interactions in water.  The work below
is a basic theoretical investigation of the success of these recent
calculations.

The developments below are based upon formal theories of association
equilibrium that have a long
history\cite{frenkel,band,hill:p_clusters,Stillinger:63}.  Formal
developments of that kind have been directed in several different
ways.  One is the study of mechanisms of
condensation\cite{Stillinger:63,fisher,swami,LaViolette:83,phillips}.
A second is towards development of theories of molecular liquids in
which the molecular species of interest are formed by combination of
atoms\cite{Pratt:76,Pratt:77}.  A third way is towards a theory of
associated liquids, like liquid water\cite{Andersen:73,Andersen:74,%
Dahl:83a,Dahl:83b,Wertheim:84a,Wertheim:84b,Wertheim:86a,Wertheim:86b,%
Wertheim:87,Wertheim:88,Stell:1996}. The latter two goals often
overlap and the formal theory is interesting more
broadly\cite{Stell:1996}.

For liquid water, the idea of distinguishable association species is a
firmly entrenched historical precedent\cite{rontgen,robinson}.  These
ideas are called ``mixture models''
\cite{eisenberg,frank,ben-naim:73,ben-naim:74,Bartell:1997}.  Though
the mixture models are common and intuitive ideas, they have never
been developed to a satisfactory theoretical conclusion.  The
available computer simulation data has always suggested that the
molecular description of liquid water is more subtle than is typically
imagined when mixture models are discussed\cite{stillinger:80}.  It
should be noted that the search continues for structural species of
special significance in the computer simulation of aqueous solutions
\cite{stillinger:80,Stillinger:acs,Stillinger-David,stillinger:83,%
corongiu,thg:95,thg:97}; but the theoretical connection of such structures to
the experimental thermodynamics requires further elaboration. One goal
of this work is to clarify the foundations of those ideas and efforts.

The theory developed below is akin to good approximations of
historical and pedagogical importance in the areas of cooperative
phenomena and phase transitions\cite{brush}.  In those areas, similar
approximations are called Guggenheim, quasi-chemical, or
Bethe approximations.  However, those treatments typically have been
developed for lattice gases, utilizing specialized considerations
appropriate for those specialized settings.  Our work here emphasizes
application to fluids, without initial lattice discretizations, and
utilizing modern tools of computational chemistry.  As one example,
some packing issues that are typically preempted by lattice gas
assumptions must be addressed explicitly here.  Thus, these
derivations and the principal results of them have not been given
before.

As a final introductory point, we note the applications of
quasi-chemical approximations in treating {\it lattice\/} models of
water\cite{Fleming:74a,Fleming:74b,Stillinger:75,Arakawa,Bell,%
Meijer:81,Meijer:82,Bodegom:84,Huckaby,Vanroyen,Sastry,Borick:93,%
Huinink:96,Besseling:94,Borick:95,Roberts,Otoole,Besseling:97}.  Those
efforts may have received less attention than they deserve because of
their lack of conventional molecular realism.  Indeed, such
calculations sometimes must make arbitrary, prior decisions that may
preclude answers to subsequent questions.  For example, if pentagonal
H-bond rings, or some other specific
structure\cite{stillinger:80,thg:95,thg:97}, were crucial for a
particular phenomenon, that issue might have to be specifically
anticipated and accomodated in typical lattice gas treatments.
However, the general success of these approaches should teach us
something about how to formulate less restricted theories of liquid
water and learning those lessons is also one of the goals of this
paper.

\section{Theory}
We consider two different clustering concepts.  The first clustering
concept is the more standard\cite{Stillinger:63,LaViolette:83}, more
global, and more ambitious.  Let lower case Greek letters identify
basic chemical components.  For each component pair $\alpha\gamma$,
provide a geometric criterion that determines whether a particular
$\alpha\gamma$ pair of particles are clustered. Clusters are then
identified in a many-body system by the rule that any pair of
particles that satisfy the pair clustering criterion are members of
the same cluster.  Despite the simplicity of this definition, it holds
a fundamental difficulty for theories of liquids: for intuitive
clustering criteria, dense liquids are typically past the percolation
threshold.  The cluster size distribution will include large clusters
that have to be directly considered.

The second clustering concept was foreshadowed by the calculations of
Hummer, {\it et al.\/} \cite{Hummer:97c} and is more local.  Focus
attention on only one particle.  Again consider a definite geometric
clustering criterion for all pairs of species types.  Then the
clusters are only those involving the distinguished particle as the
central element, or {\it nucleus\/}.  These are star-type clusters
nucleated on the central element.  For example, if the distinguished
particle is of type $\alpha$, the clusters considered are those for
which (0,1,$\ldots$) neighbors of the distinguished particle are
within the geometric $\alpha\gamma$ clustering criterion for all
$\gamma$.  The size of these clusters will be limited by the maximum
coordination number of the distinguished central particle.  This will
be a practical advantage.  But there is the corresponding
disadvantage: in the cases that particular extended clusters are
expected on a physical grounds to be especially significant, those
extended clusters may not have a direct role to play in this theory.

Theories developed for these different clustering concepts will
eventually diverge from each other.  But they can be characterized by
equations of similar form in which cluster properties play a decisive
role.  The derivation of these equations is our goal.

\subsection{Preliminaries} \noindent 
Here we make some preliminary comments that serve to simplify the
subsequent derivations and present some of the notation used.  A
central feature of this development is the potential distribution
theorem\cite{Widom:82}.  For an atomic solute with no internal degrees
of freedom this may be expressed as:
\begin{eqnarray}\rho_\nu \Lambda_\nu^3 = \langle
e^{-\Delta U /RT}\rangle _0 \ z_\nu .\label{pdt0} \end{eqnarray} $z_\nu
= e^{\mu_\nu /RT}$ is the absolute activity of the $\nu$ particle and
$ \Lambda_\nu$ is its thermal deBroglie wavelength.  The subscripted
brackets indicate the thermal average in the absence of interactions
between the solvent and the solute (test particle).  Here the averaged
property is the Boltzmann factor of the mechanical potential energy of
interaction between solvent and solute.  An equivalent description of
this bath factor is that $ \langle e^{-\Delta U /RT}\rangle _0 $ is
the average of the Boltzmann factor of the solute-solvent interactions
over the thermal motion of the solute and solvent under the condition
of no interactions between these two systems.  [We note that these
results are not limited to pairwise decomposable interactions; the
quantity $\Delta U$ is the difference between the potential energy of
the composite system and that of the separate non-interacting
systems.]  Permitting the possibility of internal degrees of freedom
including orientational degrees of freedom, the required
generalization is\cite{Widom:82}:\begin{eqnarray}\rho_\nu V/q_\nu =
\langle e^{-\Delta U /RT}\rangle _0 \ z_\nu .\label{pdt1}
\end{eqnarray} $q_\nu \equiv q (N_\nu=1,V,T)$ is the canonical
partition function for the system of one molecule of type $\nu$ in
volume V at temperature T.

Fundamental results below that are central to our derivation can be
viewed as formally exact generalizations of this potential
distribution expression for the case that molecular clusters form.
For those purposes, we will require some elaborations of notation.  We
suppose that a geometric criterion has been given by which a cluster
of type M is recognized and that this criterion is expressed by an
indicator function $H_M$; $H_M = 1$ when a cluster of type M is formed
and zero when it is not.  An ``M-clustered'' configuration is one for
which $H_M = 1$.  The results below will involve the canonical
partition function, $q_M$ for a cluster of type M; this is understood
to be the partition function of the particles that compose a cluster
of type M over the region $H_M = 1$.  Suppose that an M-clustered
configuration is given and consider placements of particles other than
those that are M-clustered.  Not all configurations of these
additional particles can be permitted without contradiction of the
specification that a particular M-cluster is present.  A further
extension of this notation will use $H_{N\vert M} = 1$ to indicate
that region wherein the N-M other particles in the N-body system are
{\it outside\/} the clustering condition for an M-cluster.  $H_{N\vert
M} = 0$ for positions of those additional species that are not
permitted under the condition that the initially specified particles
are M-clustered.  We then consider bath factors denoted by $\langle
e^{-\Delta U /RT}H_{N\vert M} \rangle _{0,M}$.  This will indicate the
average over the thermal motion of the M-cluster and the solvent under
condition of no interaction between them.  The averaged quantity
involves the exclusion factor $ H_{N\vert M}$ in addition to the
familiar Boltzmann factor of the solute-solvent interactions.  This
essential exclusion factor then assigns the value {\it zero\/} as the
weight for those configurations for which the solvent penetrates the
M-clustering volume.

\subsection{Global Clustering Definition} \noindent 
In order to involve information on clusters, we express the density of
interest in terms of cluster concentrations $\rho_M$. Thus, for the
density of the $\alpha$ particles, we would write
\begin{eqnarray}\rho_\alpha = \sum_M  n_{\alpha M}\rho_M
\label{stoichiometry} \end{eqnarray} where M identifies a molecular
cluster considered, $n_{\alpha M}$ is the number of $\alpha$ species
in that cluster, and the sum is over all molecular clusters that can
form.

The cluster concentrations $\rho_M$ are obtained from \begin{eqnarray}
\rho_M = (q_M/V)\langle e^{-\Delta U /RT}H_{N\vert M} \rangle _{0,M}
\prod_\gamma z_\gamma^{n_{\gamma M}} . \label{pdt}
\end{eqnarray} $q_M=q_M(T,V)$ is the canonical partition function
covering  configurations of an M-cluster
\cite{Stillinger:63,LaViolette:83,Pratt:76,Pratt:77}.  The indicated
average utilizes the thermal distribution of cluster and solvent under
the conditions that there is no interaction between them.  $\Delta U$
is the potential energy of interaction between the cluster and the
solvent.

Eq.~\ref{pdt} can be derived by considering the grand
ensemble.  The average number $ <N_M>$ of such clusters is composed as
\begin{eqnarray} \Xi({\bf z},T,V)<N_M> &  = &   \prod_\sigma
z_\sigma^{n_{\sigma M}} \\ \nonumber & \times & \sum_{{\bf N}\ge {\bf
n}_M} Q({\bf N},V,T \vert {\bf n}_M) \prod_\gamma { N_\gamma \choose
n_{\gamma M} } z_\gamma^{N_\gamma-n_{\gamma M}} . \label{derive1}
\end{eqnarray} Here $\Xi({\bf z},T,V) $ is the grand canonical
partition function; $N_\gamma$ is the total number of $\gamma$
particles in the system; {\bf N} is the set of particle numbers
\{N$_\gamma \cdots$\} and similarly {\bf n}$_M$ is the set of particle
numbers for the M-cluster, \{n$_{\gamma M},\cdots\}$; $Q({\bf N},V,T
\vert {\bf n}_M) $ is the canonical ensemble partition function with
the constraint that {\bf n}$_M$ specific particles are clustered.
This constraint means that the general integrand range has been
partitioned and this integration is weighted by $H_M H_{N|M}$ for
specific {\bf n}$_M$ particles clustered.  The binomial coefficient
${N_\gamma\choose n_{\gamma M}}$ is the number of $n_{\gamma
M}$-tuples of $\gamma$ particles that can be selected from $N_\gamma$
particles.  Because of the particle number factors in the summand the
partition function there can also be considered to be the partition
function for {\bf N}-{\bf n}$_M$ particles but with an extra,
distinguished {\bf n}$_M$ objects that constitute the cluster of
interest.  A natural distribution of those {\bf n}$_M$ extraneous
objects is the distribution they would have in an ideal gas phase; the
Boltzmann factor for that distribution appears already in the
integrand of the $Q({\bf N},V,T \vert {\bf n}_M) $ and the normalizing
denominator for that distribution is $ q_M(T)\prod_\gamma n_{\gamma
M}! $. The acquired factorials cancel the denominators of the binomial
coefficients.  The remaining fragments of the binomial coefficients
merely adjust the factorials involved in the definition of $Q({\bf
N},V,T \vert {\bf n}_M)$.  Final alignment of the dummy summation
variables then leads to Eq.~\ref{pdt}.

Combining our preceding results, we obtain \begin{eqnarray}
\rho_\alpha = \sum_M n_{\alpha M} ( q_M / V ) \langle e^{-\Delta U
/RT}H_{N\vert M} \rangle _{0,M} \prod_\gamma z_\gamma^{n_{\gamma M}} .
\label{answer1}\end{eqnarray} This is an equation that might be solved
for the absolute activities {\bf z}=(z$_\alpha, \ldots$) in terms of
the densities, cluster partition functions, and the temperature.  If
the bath contribution is neglected, $\langle e^{-\Delta U
/RT}H_{N\vert M} \rangle _{0,M} $=1, this is exactly the relation of
Eq.~11 of \cite{LaViolette:83}. This is the result that was sought.

Though formally correct, there is a fundamental difficulty with this
result.  Consider a clustering definition in which particles at
near-neighbor distances in solution are clustered.  Then the sum will
diverge as a percolation threshold is approached by increasing the
density of a dilute phase.  This is true whether or not cluster
interference is neglected.  This divergence is sometimes taken as a
practical indication of condensation at low temperatures.  However,
the sum will diverge at similar densities even if no condensation
occurs.  Thus, this formula is inapplicable to a liquid, traditionally
defined, without further considerations.

Notice however that there is a non-trivial special case for which
Eq.~\ref{answer1} may be directly applied to a condensed phase and is
likely to be helpful.  This is the case where species $\alpha$ is a
dilute solute and the interest is in the effect of the solvent on
$\mu_\alpha$.  Then we may adopt such a restricted definition that no
solvent-solvent clusters can form.  However, at the same time we can
define the solute-solvent clustering criteria more physically and
study those clusters in which solvent molecules bind to the solute of
interest.  Those clusters will be finite and the sum of
Eq.~\ref{answer1} will involve only a finite number of terms.

\subsection{Local Clustering Definition} \noindent
We will derive the result needed through an indirect argument that
utilizes the already derived Eq.~\ref{answer1}.  We wish to consider
non-dilute systems and species for which Eq.~\ref{answer1} does not
apply directly.  We will find a way to use  Eq.~\ref{answer1} by
appropriately distinguishing single molecules in this non-dilute
phase.

We begin by noting the well-known fact that the chemical potential,
say $\mu_\alpha$, can be divided into ideal and interaction parts.
The ideal contribution takes the form $RT\ln \rho_\alpha + constant$
where the constant might be calculated on the basis of molecular
properties; see Eq.~(\ref{pdt0}).  The first step in our argument is
the specification that the theory need only determine the interaction
part of the chemical potential since the ideal contribution is
well-known.  To determine the interaction part of the chemical
potential we distinguish a single molecule of type $\alpha$ and study
its condition in solution.  This is natural; for example a simulation
calculation might select a particular $\alpha$ molecule and perform
charging or uncharging calculations, or determine distributions of
binding energies experienced\cite{Hummer:97c}.  When an $\alpha$
molecule is selected for the purposes of calculation of the
interaction part of $\mu_\alpha$ it can be treated as a solute at the
lowest non-zero concentration, as a solitary impurity.  We will denote
the chemical potential of this distinguished solute as
$\mu_{\alpha'}$, remembering that the interaction part of
$\mu_{\alpha'}$ will be the same as the interaction part of
$\mu_{\alpha}$.  For a dilute solute, we can define clustering
criteria, as anticipated above, so that no solvent-solvent clustering
occurs as defined, but the definition of clustering of solvent
molecules about the distinguished $\alpha$ solute is naturally
included.  Eq.~\ref{answer1} then does apply to the calculation of
$\mu_{\alpha'}$.

The modifications of Eq.~\ref{answer1} for this case are two: First,
the stoichiometric coefficients of Eq.~\ref{stoichiometry}, that
appear later in Eq.~\ref{answer1}, are all one (1); since the
distinguished solute is at the lowest non-zero concentration there
cannot be more than one such solute in any cluster.  The right side of
Eq.~\ref{answer1} is precisely proportional to z$_{\alpha'}$.  Second,
all clusters are of star type, that is, AB$_n$ with the distinguished
solute at the center.

Before finally writing the desired result, we ask again about the
ideal contribution to $\mu_{\alpha'}$ and to $\mu_{\alpha}$.  This
ideal contribution is reflected in the density on the left of
Eq.~\ref{answer1}.  For $\mu_{\alpha}$ that density is the physical
value and is part of the definition of the problem.  For
$\mu_{\alpha'}$, if our argument were taken literally, that density on
the left would be $\rho_{\alpha'} = 1/V$.  Replacement of that value
by $\rho_\alpha$, and on the right simultaneously z$_{\alpha'}$ by
z$_{\alpha}$, would merely readjust $\mu_{\alpha'}$ up to
$\mu_{\alpha}$ through a final assessment of the ideal contribution.
Thus, we have \begin{eqnarray} \rho_\alpha = \sum_{M(\alpha)} ( q_M /
V ) \langle e^{-\Delta U/RT}H_{N\vert M} \rangle _{0,M} \prod_\gamma
z_\gamma^{n_{\gamma M}} .\label{answer2}\end{eqnarray} The sum is over
all clusters M($\alpha$) that can form on an $\alpha$ nucleus.

A practical example of the importance of the clustering definition may
be helpful.  Recent work on clusters of a chloride ion with water has
suggested that the preferred disposition of the chloride ion may be
near the surface of the cluster\cite{Perera:93}.  This interesting
point is unlikely to be decisive for the application of this cluster
formula to the study of the solvation of the chloride ion in liquid
water.  The cluster definitions here require that the chloride be the
{\it nucleus\/} of a star-type cluster. That would permit the chloride
ion to access the physical surface of a droplet only for small
clusters and larger clusters are likely to be decisive in establishing
the bulk phase thermodynamics of the aqueous solutions containing
chloride ions.

\subsection{Cluster Interference} \noindent
This paper will develop the local clustering alternative; the global
clustering results will serve as contrast.  The issue of cluster
interference is different in the two cases.  The development requires
a complete and unique partitioning of phase space into regions
characterized by a specific cluster population.  For each proper
configuration, the definition uniquely assigns elementary particles to
clusters.  With this characteristic, we then formally regard the
cluster populations as supplementary integration (summation)
variables, first integrating over configurations with a specific
constraint of a specific cluster population, then summing over
permitted cluster populations.  Cluster interference is a simple
implementation of the constraint.  If a particular cluster of type M
is specified then configurations that violate the specification cannot
be allowed.  As a particular example, suppose that an A$_n$ cluster is
under consideration.  Then some configurations for which an additional
A particle approaches the A$_n$ cluster must be excluded; otherwise
configurations of A$_{n+1}$ clusters would become confused with
configurations of A$_n$ clusters.  In the notation above these
constraints are lumped into the factors $\langle e^{-\Delta
U/RT}H_{N\vert M} \rangle _{0,M}$ through rigid exclusion
interactions.  It is in these bath factors and the cluster partition
functions $q_M$ that cluster interference is expressed.

For the global clustering development, these cluster interference
contributions are complicated because they depend on all the cluster
sizes and shapes.  However, the global clustering result
Eq.~\ref{answer1} is fundamentally inapplicable to liquids.

In contrast, for the local clustering development cluster interference
is much simpler.  If we specify that a cluster of type A$_n$ is
considered, we must only require that the n-1 particles are within the
clustering volume $v$ of a distinguished molecule (easy to do) and
that no more than n-1 additional particles are there.  This latter
factor is familiar from studies of packing problems in
liquids\cite{Pratt:97b} and in the potential distribution factor it
involves the probability that the clustering volume $v$ is {\it
empty\/} of solvent species\cite{Widom:78}.  We can consider the
condition that the clustering volume is empty of solvent molecules by
introducing the probability for that event, $p_0$.  The van der Waals
approximation $p_0\approx (1-\rho v)$ should be qualitatively
satisfactory and provides definiteness to the discussion.  Thus,
within the local clustering approach, cluster interference is
completely expressed by the form \begin{eqnarray} { \rho_\alpha \over
z_\alpha } = p_0 \sum_{M(\alpha)} ( q_M / V ) \langle e^{-\Delta U/RT}
\rangle^\ast _{0,M} \prod_\gamma{}^\prime z_\gamma^{n_{\gamma M}}
.\label{answer4}\end{eqnarray} Now the well-decorated term $\langle
e^{-\Delta U/RT} \rangle^\ast _{0,M}$ indicates the average over the
thermal motion of the M-cluster and solvent under that condition that
the only interactions between them rigidly enforce the exclusion of
the solvent from the M-clustered volume.  We have factored out the
$z_\alpha$ because this quantity must be present in each term, because
the ratio $ \rho_\alpha/ z_\alpha $ is a standard form, and because
the distinguished $\alpha$ particle that is the nucleus of the cluster
requires a different treatment than do the particles on the periphery
of the star. The notation $\prod_\gamma^\prime$ means that the term
for the species nucleating the cluster should be stricken from the
product.  Though this result is formally complete, an approximate
theory will have to be utilized for  $p_0$ in specific
applications.

\subsection{Quasi-chemical Approximation}\noindent 
A theory with quasi-chemical form is \begin{eqnarray} \langle
e^{-\Delta U/RT} \rangle^\ast _{0,M} \prod_\gamma{}^\prime
z_\gamma^{n_{\gamma M}} \approx\prod_\gamma{} ^\prime \{\rho_\gamma
(V/q_\gamma)\}^{n_{\gamma M}}. \label{qca1} \end{eqnarray} This
replacement is motivated by the desire to replace the `bare fields'
$\ln z_\sigma$ with effective fields, by the recognition that
Eq.~\ref{pdt1} provides a pattern for that replacement, and by the
appreciation that the bath contributions might reasonably factor for
species on the ``surface'' of the cluster.

Note that the list of clusters M($\alpha$) should include the monomer.
One term in this list for Eq.~\ref{answer4} will have $q_M =
q_\alpha$.  Thus, we can write
\begin{eqnarray} {\rho_\alpha V \over z_\alpha q_\alpha} \approx  p_0
(1 + \sum_{M(\alpha)}{}^\prime K_M(T) \prod_\nu{}^\prime
\rho_\nu^{n_{\nu M}}) , \label{answer5} \end{eqnarray}
where
\begin{eqnarray} K_M(T) ={ q_M /V \over \prod_\nu \{ q_\nu / V
\} ^{n_{\nu M}} } \label{coefs} \end{eqnarray} 
and the sum of Eq.~\ref{answer5} is over the list of possible star
clusters nucleated by an $\alpha$ species but {\em not} including the
monomer cluster.  This theory deserves the appellation
``quasi-chemical'' because the coefficients $K_M(T)$ are the chemical
equilibrium ratios for the formation of star clusters in a dilute
gas\cite{mcq}.  Note, however, that the factor $p_0$ will be essential
for description of packing effects and thus for predictions of
thermodynamic properties of condensed phases.  The thermodynamic
quantity sought is the chemical potential
\begin{eqnarray} \mu_\alpha \approx RT\ln [{ \rho_\alpha V/ q_\alpha}
] - RT \ln [ p_0 (1 + \sum_{M(\alpha)}{}^\prime K_M(T)
\prod_\nu{}^\prime \rho_\nu^{n_{\nu M}}) ] .  \label{answer6}
\end{eqnarray}  This formula makes the conventional, helpful
separation between the contributions of intermolecular interactions
and the non-interaction (ideal) terms; see Eq.~\ref{pdt1}.  Quantities
such as $p_0$ and the $K_M(T)$ depend on parameters that define the
clustering circumstances.  But, since the physical problem is
independent of those parameters, the theory should be insensitive to
them if it is to be satisfactorily accurate.

\subsection{Discussion}\noindent  The formal results Eqs.~\ref{answer2}
and \ref{answer4}, and the approximation Eq.~\ref{answer6} are
believed to be new.  The approximation Eq.~\ref{qca1} attempts to
eliminate the complicated bath contributions.  These quantities are
formally well-understood and can be formally analyzed with Mayer
mathematical cluster expansions or functional analyses. Here we
discuss physically what's neglected.  For dilute solutions where the
solvent activities are known separately, Eq.~\ref{answer4} may be used
directly and the issues that follow here are not relevant.  This was
the pattern of the motivating calculations
\cite{Hummer:97c,Martin:97}.

Eq.~\ref{qca1} assumes that each activity factor may be replaced using
the relation Eq.~\ref{pdt1} with no account of the interference
between different sites that is formally expressed in the left-side of
Eq.~\ref{qca1}.  This might be correct for idealized circumstances,
e.g. a ``Bethe lattice'' (no cycles).  [An alternative derivation
based upon diagrammatic arguments makes it clear that this is a {\em
tree} approximation.]  But for more realistic continuous problems
there are two sources of that interference.  Firstly, different
peripheral sites on the star cluster can interfere with one another.
This effect arises because solvent molecules can interact with two or
more surface sites jointly.  The organization of the problem here is
designed to mitigate that interference between different surface
species.  Secondly, ``through solvent'' interference between a
peripheral site and the nucleus of the cluster arises when solvent
molecules can interact with a surface surface sites and the nucleus or
cluster volume jointly.  Eq.~\ref{qca1} neglects both of these
effects.

We reiterate that the quantities neglected by the approximation
Eq.~\ref{qca1} are well understood formally.  Thus, in specific
applications it should be practical to augment this approximation with
additional contributions that are understood physically and
theoretically.  An example should serve to clarify this point.  There
has been significant recent interest in primitive electrolyte solution
models under circumstances where ion pairing and clustering is
acknowledged to be of primary significance\cite{zuckerman}.  The
formal developments here apply also to ion clustering in electrolyte
solutions.  However, where non-neutral clusters are important the
approximation Eq.~\ref{qca1} must be at the least augmented to treat
long range interactions, likely following a random phase
approximation.

A more specific example is given by the study of the hexa-aquo ferric
ion, Fe(H$_2$O)$_6$$^{3+}$ reported in Reference \cite{Martin:97}.
The data given there allow us to estimate the error of the neglect of
long-ranged interactions, Eq.~\ref{qca1}.  That neglected contribution
would be about 391 kcal/mol, or 38\% of the value inferred from
experiment for interaction part of $\mu_{Fe^{3+}}$.  [See Table IV of
that report but note that the packing contribution here present as
$p_0$ was neglected in that previous study.]  Thus, for ionic solutes
in particular, further consideration of coulomb ranged interactions
will be necessary.  Though the physical estimate given for the
hexa-aquo ferric ion example \cite{Martin:97} was natural and
surprisingly accurate, the value 391 kcal/mol was essentially
empirically derived.  To do better than that, the present approach
must be extended to produce the dielectric constant of the liquid in
order to assess screening of electrostatic interactions.  Since the
this approach has a conceptual overlap with the Onsager-Kirkwood
\cite{Kirkwood,OK} development of the theory of dielectrics that
subsequent step should be natural.  We note, however, that the present
hybrid approach will improve the treatment of negative ions in
solution, such as the Cl$^-$ ion which was a remaining difficulty for
the multistate gaussian model \cite{Hummer:97c}.

Finally, the hexa-aquo ferric ion example emphasizes that current
electronic structure software can routinely produce $K_M(T)$ in a
harmonic approximation.  Other recent examples include the work of
references \cite{weinhold,lwf:98a,lwf:98b,lwf:98c}.  This encourages
us to anticipate that further developments will permit implementation
of these theories without an intermediate effort to obtain approximate
force field models.  In the near term, this approach should at the
least offer a direction for improvement of electronic structure
calculations on solution species, calculations that might presently
rely solely on dielectric continuum models.

\section{Conclusions}
The traditional development of theories of fluids based upon the
concept of physical clustering can be adapted to an alternative local
clustering definition.  The alternative clustering definition can
straightforwardly preserve a detailed valence description of the
interactions between a solution species and its near-neighbors, {\it
i.e.,\/} cooperativity and saturation of coordination for strong
association.  These clusters remain finite even for condensed phases.
The simplest theory to which these developments lead is analogous to
quasi-chemical theories of cooperative phenomena.  The present
quasi-chemical theories require additional consideration of packing
issues because they don't impose lattice discretizations on the
continuous problem.  These quasi-chemical theories do not require pair
decomposable interaction potential energy models.  Since calculations
may be required only for moderately sized clusters, we anticipated
that these quasi-chemical theories could be implemented with
computational tools of current molecular electronic structure theory.

\section*{Acknowledgement}  This work was supported by the LDRD
program at Los Alamos.  The work at INEEL was performed under DOE Idaho
Operations Office Contract DE-AC07-94ID13223.

\end{document}